\documentclass{WileyMSP-template}
\usepackage{amsmath,amssymb,amsfonts}
\usepackage[colorlinks,urlcolor=blue,linkcolor=blue,citecolor=blue,anchorcolor=blue]{hyperref}
\usepackage{cite}
\usepackage{palatino}
\usepackage{esint}
\usepackage[numbers,sort&compress]{natbib}

\begin{document}

	\pagestyle{fancy}
	\rhead{\includegraphics[width=2.5cm]{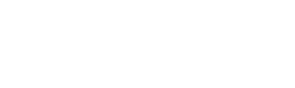}}
	
	\title{\noindent Observation of nonlinear topological corner states originating from different spectral charges}
	
	\maketitle
	
	\noindent
	\author{Victor O. Kompanets$^{1,\dagger}$, Suge Feng$^{2,\dagger}$, Yiqi Zhang$^{2,\dagger,*}$, Yaroslav V. Kartashov$^{1,*}$, Yongdong Li$^2$, Sergei A. Zhuravitskii$^{1,3}$, Nikolay N. Skryabin$^{1,3}$, Alexander V. Kireev$^{1}$, Ivan V. Dyakonov$^{3}$, Alexander A. Kalinkin$^{1,3}$, Ce Shang$^4$, Sergei P. Kulik$^3$, Sergey V. Chekalin$^{1}$, Victor N. Zadkov$^{1,4}$}

	\begin{affiliations}	\noindent
		$^1$ Institute of Spectroscopy, Russian Academy of Sciences, Troitsk, Moscow 108840, Russia\\
		$^2$ Key Laboratory for Physical Electronics and Devices, Ministry of Education, School of Electronic Science and Engineering, Xi'an Jiaotong University, Xi'an 710049, China \\
		$^3$ Quantum Technology Centre, Faculty of Physics, M. V. Lomonosov Moscow State University, Moscow 119991, Russia\\
		$^4$ Aerospace Information Research Institute, Chinese Academy of Sciences, Beijing 100094, China\\
		$^5$ Faculty of Physics, Higher School of Economics, Moscow 105066, Russia \\
		$^\dagger$ These authors contributed equally to this work.	\\
		$^*$ Email Address: \href{mailto: zhangyiqi@xjtu.edu.cn}{zhangyiqi@xjtu.edu.cn}, \href{mailto: kartashov@isan.troitsk.ru}{kartashov@isan.troitsk.ru}\\
	\end{affiliations}

	\noindent

	\begin{abstract} \noindent
		Higher-order topological insulators (HOTIs) are unique topological materials supporting edge states with the dimensionality at least by two lower than the dimensionality of the underlying structure. HOTIs were observed on lattices with different symmetries, but only in geometries, where truncation of HOTI produces a finite structure with the same order of discrete rotational symmetry as that of the unit cell, thereby setting the geometry of insulator edge. Here we experimentally demonstrate a new type of two-dimensional (2D) HOTI based on the Kekul\'e-patterned lattice, whose order of discrete rotational symmetry differs from that of the unit cells of the constituent honeycomb lattice,  with hybrid boundaries that help to produce all three possible corners that support effectively 0D corner states of topological origin, especially the one associated with spectral charge $5/6$. We also show that linear corner states give rise to rich families of stable hybrid nonlinear corner states bifurcating from them in the presence of focusing nonlinearity of the material. Such new types of nonlinear corner states are observed in hybrid HOTI inscribed in transparent nonlinear dielectric using fs-laser writing technique. Our results complete the class of HOTIs and open the way to observation of topological states with new internal structure and symmetry.
	\end{abstract}
	

	\section{Introduction}
	
	Photonic topological insulators~\cite{lu.np.8.821.2014, ozawa.rmp.91.015006.2019, zhang.nature.618.687.2023, lin.nrp.5.483.2023} representing the generalization of electronic topological insulators~\cite{hasan.rmp.82.3045.2010, qi.rmp.83.1057.2011} to the realm of electromagnetic systems, are principally novel materials, where light waves demonstrate the unique propagation properties and where new possibilities for molding the flow of light, its switching and routing arise. Main advantages of topological insulators are related to exceptional robustness of the in-gap edge states appearing at the edges of such structures and stemming from their protection by the very topology of the system that cannot be changed without closing the forbidden topological gap. This protection allows edge states to survive in the presence of disorder, and bypass edge defects or corners without radiation or reflection --- the properties that are very attractive for potential design of topologically protected optical transmission and routing devices. Different types of photonic topological insulators have been proposed that are tightly connected with particular mechanism of the appearance of topologically nontrivial phase, should it be breakup of certain symmetries of the system~\cite{wang.nature.461.772.2009, noh.prl.120.063902.2018, xia.prl.131.013804.2023, xie.acs.11.772.2024}, controllable deformations of the unit cells of underlying lattice structure~\cite{banalcazar.science.357.61.2017, benalcazar.prb.96.245115.2017, song.prl.119.246402.2017, schindler.sa.4.aat0346.2018, xie.nrp.3.520.2021}, or modulations of parameters of the system in evolution variable~\cite{rechtsman.nature.496.196.2013, maczewsky.nc.8.13756.2017, mukherjee.nc.8.13918.2017, pyrialakos.nm.21.634.2022}, to name a few. The common property of all such systems is that the symmetry of the underlying lattice forming topological insulator and geometry of its edge are important factors determining the properties and domains of existence of topological edge states.
	
	This is also the case for recently discovered intriguing class of higher-order topological insulators (HOTIs) known for lower-dimensional edge states that they can support (in such insulators the difference between dimensionality of the system and effective dimensionality of some edge states can be two or even higher)~\cite{xie.nrp.3.520.2021}. For example, a two-dimensional (2D) HOTI can support 0D corner states. As a new kind of topological matter, HOTIs can be divided into two general categories: the insulators that are characterized by the nonzero quantized multipole moments~\cite{banalcazar.science.357.61.2017, serra.nature.555.342.2018, ni.nc.11.2108.2020, xue.nc.11.2442.2020}, and HOTIs for which such multipole moments are absent~\cite{benalcazar.prb.99.245151.2019, xie.prl.122.233903.2019, chen.prl.122.233902.2019, li.np.14.89.2020, kirsch.np.17.995.2021, shang.as.11.2303222.2024}. The latter HOTIs are also frequently called higher-order crystalline topological insulators~\cite{wieder.nrm.7.196.2022}, with their band topology being protected by the discrete rotational symmetry of the underlying lattice~\cite{peterson.nature.589.376.2021, liu.nature.589.381.2021}. In contrast to known types of the first-order topological insulators~\cite{rechtsman.nature.496.196.2013, klembt.nature.562.552.2018, noh.prl.120.063902.2018}, HOTIs do not obey the bulk-edge correspondence principle~\cite{ozawa.rmp.91.015006.2019}, while topological phase transition in HOTIs can be characterized by the polarization index invariant~\cite{xie.nrp.3.520.2021}. In all cases where lower-dimensional edge states appear in HOTI, should it be corner states emerging in the outer corners of the truncated lattice or disclination states appearing around the inner topological defects, their appearance can be associated with the ``filling anomaly'' --- in language of solid-state physics, a mismatch that appears between the number of electrons in an energy band and the number of electrons required for charge neutrality~\cite{benalcazar.prb.99.245151.2019, li.prb.101.115115.2020, peterson.science.368.1114.2020, peterson.nature.589.376.2021}. This anomaly indicates on the appearance of so-called fractional charges in the corners of the structure. It has been shown that the analogue of the above mentioned charge in optical systems is the ``spectral charge'' characterizing the ``number of modes'' or ``occupied sites'' in the unit cell of the lattice that can be calculated using all states below the bandgap~\cite{liu.nature.589.381.2021}. Thus, making a summation of all states (normalized field modulus) below the considered bandgap, one may find that resulting sum distributes uniformly on all sites in the bulk unit cells (i.e., all sites in the bulk unit cells are occupied), but on only few sites in the boundary unit cells with other sites being vacant.
	
	The intrinsic connection between ``filling anomaly'' and corner states in HOTI illustrates that the geometry of HOTI is exceptionally important and defines not only the symmetry of the corner states, but the very fact of their existence. Nevertheless, so far almost all experimentally observed HOTIs were constructed with the same discrete rotational symmetry as the symmetry of the unit cell of the respective lattice, that dictates the geometry of the edge and types of corners that HOTI may have. For example, this is the case for insulators observed on honeycomb lattices~\cite{noh.np.12.408.2018}, on square 2D generalizations of Su-Schrieffer-Heeger chains~\cite{peterson.nature.555.346.2018, serra.nature.555.342.2018, xie.prl.122.233903.2019, chen.prl.122.233902.2019}, and Kagome lattices~\cite{ezawa.prl.120.026801.2018, kirsch.np.17.995.2021}. In this article, using honeycomb lattice (i.e. so-called photonic graphene~\cite{plotnik.nm.13.57.2014}) that realizes a variant of Kekul\'e-patterned lattice that is frequently also termed Wu-Hu model after seminal paper~\cite{wu.prl.114.223901.2015} that proposed this type of lattice distortion for realization of the topological material possessing the spin Hall effect, we create a new type of crystalline HOTI with \textit{hybrid edge} and discrete $\mathcal{C}_3$ rotational symmetry that is lower than $\mathcal{C}_6$ discrete rotational symmetry of the unit cells of underlying lattice. The transition between topologically trivial and nontrivial regimes is achieved by adjusting the separation among the six sites in the unit cell. We show that hybrid edge of this structure can be designed such as to contain all possible corners, each of which can support effectively 0D topological corner states, whose appearance is connected with fractional spectral charges $1/3$, $1/2$, and $5/6$ that can be evaluated using the Wannier center method. Notice that in our system hybrid corner states coexist with, but have symmetry distinct from conventional topological states forming in the outer corners of the structure. While in small-size system hybrid corner states may interact, with increase of the system size they gradually turn into isolated 0D modes. Notice that corner states connected with different spectral charges do not require for their realization exclusively the $\mathcal{C}_3$-symmetric array configuration adopted in this work and may be potentially observed in other structures. For instance, in the \textbf{Supporting Information}, we display the spectrum of a $\mathcal{C}_6$-symmetric configuration with \textit{hybrid edges}, which also supports a variety of corner states originating from filling anomaly with different fractional charges. We stress, however, that $\mathcal{C}_3$-symmetric configuration is advantageous because unit cells with different fractional charges are well-separated in it allowing independent excitation of different types of corner states by selective focusing into different sites. It should be also mentioned that different fractional charges can be also encountered in a $\mathcal{C}_3$-symmetric configuration with pure armchair boundaries theoretically suggested in \cite{ezawa.prb.73.045432.2006, ezawa.prb.76.245415.2007}, see its representative spectrum in \textbf{Supporting Information}.
	
	We observe the aforementioned corner states in fs-laser written waveguides arrays in fused silica, the material that possesses focusing Kerr nonlinearity. The interplay between nonlinearity and topology~\cite{smirnova.apr.7.021306.2020, szameit.np.20.905.2024}, already allowed to predict and observe in experiments a number of intriguing phenomena, which include, but are not limited to the formation of topological solitons~\cite{lumer.prl.111.243905.2013, ablowitz.pra.96.043868.2017, leykam.prl.117.143901.2016, ivanov.acs.7.735.2020, mukherjee.science.368.856.2020,mukherjee.prx.11.041057.2021, zhong.ap.3.056001.2021,hu.light.10.164.2021,xia.light.9.147.2020,ren.light.12.194.2023,arkhipova.sb.68.2017.2023,zhong.light.13.264.2024}, 
	topologically protected non-Hermitian phase transitions~\cite{dai.np.10.101.2024}, nonlinearity-induced topological phases~\cite{maczewsky.science.370.701.2020, sone.np.20.1164.2024}, and high-harmonic generation in edge states~\cite{smirnova.prl.123.103901.2019}. Here we illustrate a new feature of HOTI with hybrid edge --- coexistence of several types of thresholdless nonlinear corner states appearing around different corners at the hybrid edge, that differ from previously reported nonlinear topological corner states in outer corners. We illustrate the shape and localization tunability that nonlinearity introduces into hybrid corner states. These results not only provide a new route for control of light propagation in topological systems, but also introduce a promising material platform for observation of the higher-order band topology. The first experimental demonstration of HOTIs with hybrid edges presented here leads to new insights and illustrates that the type of lattice truncation in such systems is of considerable importance for the nature, location, spatial structure and symmetry of the emerging localized modes and nonlinear states bifurcating from them. The formation of different coexisting types of topological corner states in one lattice configuration is important for practical applications and can be used for topologically-protected information transfer~\cite{peterson.nature.555.346.2018, jiang.rip.13.100112.2025}, design of new types of topological lasers~\cite{ota.nano.9.547.2020}, and design of quantum memory~\cite{heshami.jmo.63.2005.2016, moiseev.prl.134.070803.2025}.
	
	\section{Results}
	
	\subsection{HOTI configuration, spectrum and corner states}
	
\begin{figure*}[t]
		\centering
		\includegraphics[width=1\textwidth]{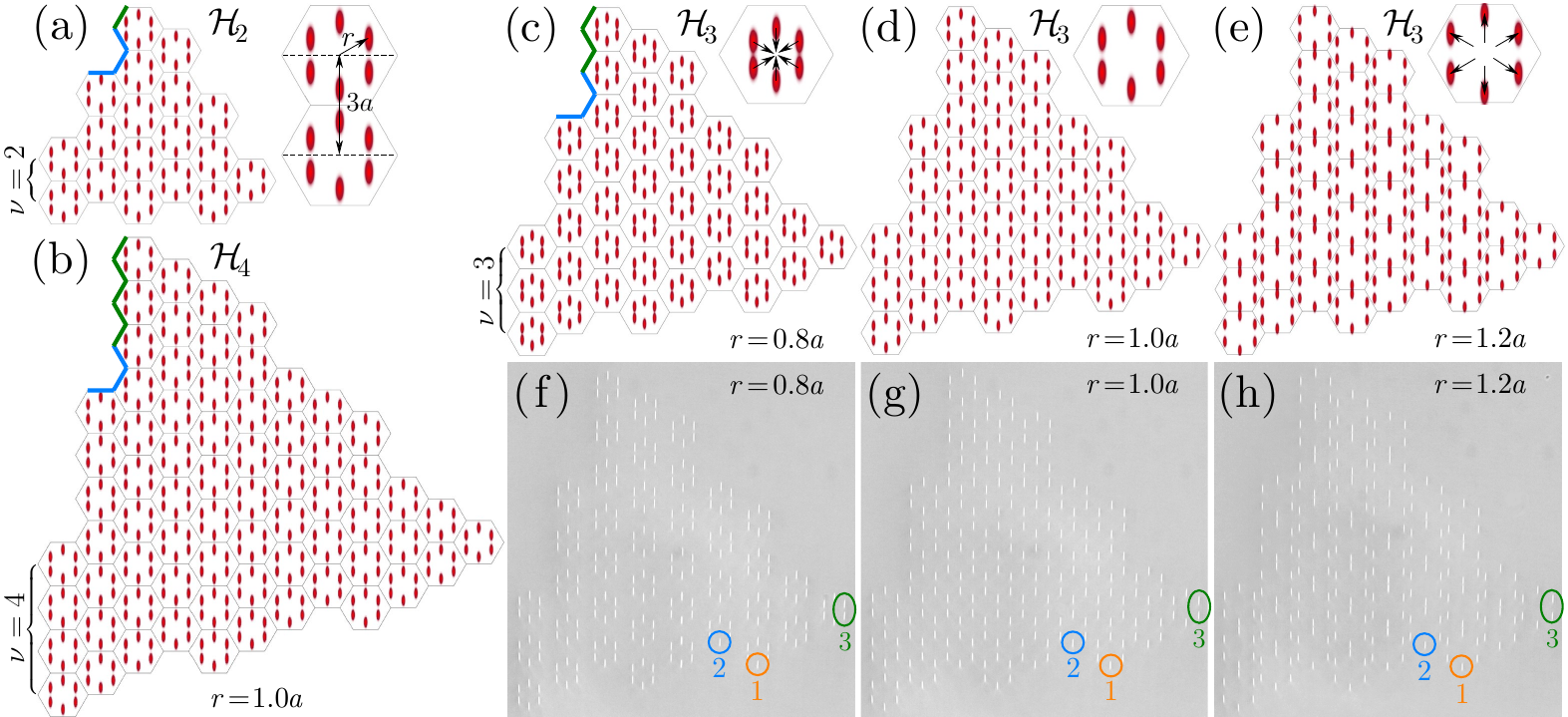}
		\caption{\textbf{Configuration of the novel higher-order topological insulators}.
			Schematic representations of ${\mathcal H}_2$ (\textbf{a}), ${\mathcal H}_3$ (\textbf{d}), and ${\mathcal H}_4$ (\textbf{b}) arrays with hybrid edges for spacing ${r=1.0a}$ between waveguides in the unit cells that are indicated by gray hexagons. The inset in (\textbf{a}) illustrates spatial parameters $a$ and $r$ that determine size of the unit cell and allow to change topology, respectively. Panels (\textbf{c})-(\textbf{e}) show transformation of the ${\mathcal H}_3$ array due to the shift of waveguides in each unit cell towards the center of the cell (\textbf{c}) and towards its edges (\textbf{e}). Insets show unit cells with arrows indicating the direction of waveguide shift, while values of intracell spacing are indicated below schematic images. (\textbf{f})-(\textbf{g}) Microphotographs of laser-written ${\mathcal H}_3$ waveguide arrays corresponding to panels (\textbf{c})-(\textbf{e}).
			The colored circles with numbers indicate the three corners that may support corner states.
			The value of $\nu$ gives the number of stacked unit cells in the left "column" of the structure. The green/blue lines in (a-c) highlight the zigzag/armchair edges.}
		\label{fig1}
	\end{figure*}

	To construct HOTI with hybrid edges we employ honeycomb unit cells, each of which contains six waveguides, as shown in the inset of Fig.~\ref{fig1}(a). 
	The constructed finite structure has $\mathcal{C}_3$ discrete rotational symmetry (if we neglect the ellipticity of the waveguides arising due to the writing procedure, which is non-essential for the results described below, but which nevertheless is taken into account), and its size is characterized by the number of cells $\nu$ stacked at the very left edge. 
	The construction procedure is shown in the \textbf{Supporting Information}. The smallest structure obtained for ${\nu=2}$ that we term $\mathcal{H}_2$ is depicted in Fig.~\ref{fig1}(a), while its larger $\mathcal{H}_3$ ${(\nu=3)}$ and $\mathcal{H}_4$ ${(\nu=4)}$ counterparts are depicted in Figs.~\ref{fig1}(c)-\ref{fig1}(e) and Fig.~\ref{fig1}(b), respectively. To control topology of the system we tune the position of waveguides in each unit cell by varying distance $r$ between waveguides and the center of the unit cell with fixed side length $a$. For ${r=1.0a}$ the spacing between neighboring waveguides in the entire lattice is identical [see Figs.~\ref{fig1}(b) and \ref{fig1}(d)], for ${r<1.0a}$ the intracell spacing between waveguides is smaller than the intercell one [Fig.~\ref{fig1}(c)], while for ${r>1.0a}$ the intracell spacing becomes larger than the intercell one [Fig.~\ref{fig1}(e)]. Such waveguide arrays can be inscribed using fs-laser writing technique (see \textbf{Materials and Methods})~\cite{rechtsman.nature.496.196.2013,maczewsky.nc.8.13756.2017,mukherjee.nc.8.13918.2017,noh.np.12.408.2018,noh.prl.120.063902.2018,wang.am.31.1905624.2019,xu.np.15.703.2021,chang.am.34.2110044.2022,chang.am.36.2310010.2024,yan.npjn.1.40.2024}. We have chosen sufficiently large $\mathcal{H}_3$ structures as a main platform that ensures excellent reproducibility of results, and we have checked that they remain valid for even larger lattices. Microphotographs of laser-written $\mathcal{H}_3$ structures for various shift parameters $r$ are presented in Figs.~\ref{fig1}(f)-\ref{fig1}(h). Here, at one edge of the sample, we indicate the three corners with colored circles that may support corner states. Notice that in contrast to previously studied HOTIs~\cite{xie.nrp.3.520.2021, lin.nrp.5.483.2023}, our lattices possess \textit{hybrid edges} containing alternating armchair and zigzag segments which are highlighted by blue and green lines, respectively, in Figs.~\ref{fig1}(a)-\ref{fig1}(c) showing structures of different orders, that results in the appearance of multiple ``corners'' along each edge of the structure. Their presence is manifested in peculiar structure of the linear spectrum of this waveguide system.
	
	The paraxial propagation of light beams in lattice with hybrid edges can be accurately described by the nonlinear Schr\"odinger equation for the dimensionless amplitude of the light field $\psi$:
	\begin{align}\label{eq1}
		i \frac{\partial \psi}{\partial z} = -\frac{1}{2} \nabla^2 \psi -\mathcal{R}(x,y) \psi-|\psi|^{2} \psi.
	\end{align}
	Here ${\nabla^2\equiv\partial_x^2+\partial_y^2}$ is the transverse Laplacian. 
	We assume that the medium is focusing, the transverse $(x,y)$ and the longitudinal $z$ coordinates are dimensionless, the function $\mathcal{R} (x,y)$ describes shallow optical potential defining hybrid lattice 
	\[
	{\mathcal{R} (x,y) =  p \sum_{m,n}  e^{- (x-x_{m,n})^2/d_x^2 - (y-y_{m,n})^2/d_y^2} },
	\] 
	where $p$ is the depth of potential proportional to the depth of the refractive index modulation, $(x_{m,n},y_{m,n})$ are the coordinates of the waveguides that vary in accordance with shift parameter $r$, the side length of the unit cell is given by $a$, $d_{x,y}$ are the waveguide widths (here we assume that ${d_y>d_x}$ to account for ellipticity of waveguides due to writing technology). Further we set ${p=4.7}$, ${a=3.3}$, ${d_x=0.25}$ and ${d_y=0.75}$ in accordance with parameters of our laser-written structures (the details of all normalization parameters are presented in \textbf{Materials and Methods}). It should be mentioned that the continuous model (\ref{eq1}) that accounts for all details of the refractive index landscape, i.e. real profile and exact locations of all waveguides in the lattice, actually takes into account all possible couplings between all waveguides in the array, thereby providing more accurate description in comparison with tight-binding models that usually consider only nearest-neighbor couplings. In addition, the model (\ref{eq1}) does describe variation of the fields inside the waveguides, that may be especially important for stability analysis of nonlinear states.
	
\begin{figure*}[t]
		\centering
		\includegraphics[width=\textwidth]{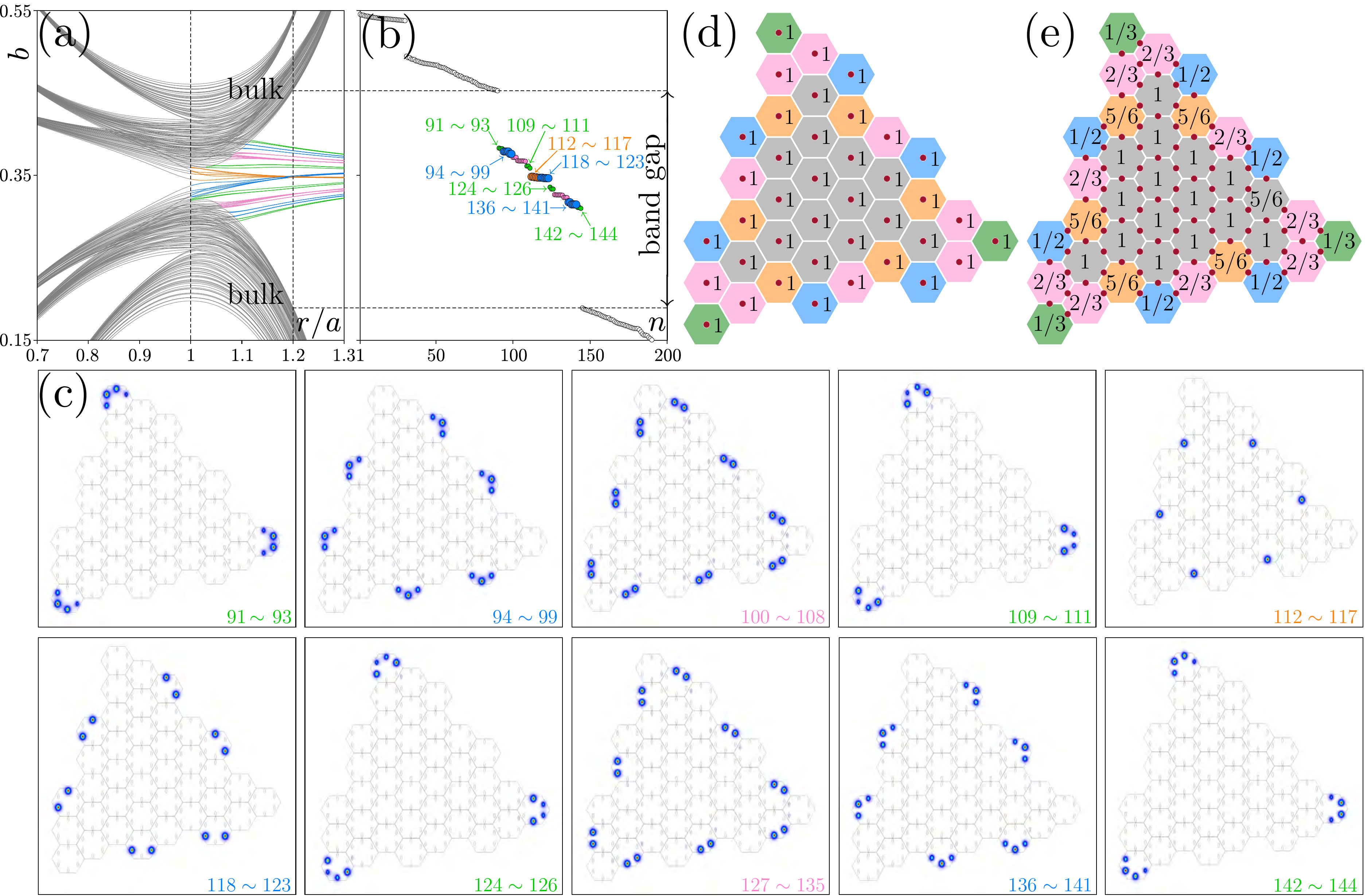}
		\caption{\textbf{Spectrum, corner states, and spectral charge of each unit cell of the ${\mathcal H}_3$ lattice.}
			(\textbf{a}) Linear spectrum showing dependencies of all eigenvalues of lattice modes versus shift parameter $r$ at ${a=3.3}$. There is a band gap between top and bottom bands of the bulk states, where various corner states appear. (\textbf{b}) Linear spectrum of the lattice at ${r=1.2a}$. (\textbf{c}) Intensity distributions in linear corner states with different indices given in the right bottom corner of each panel. Both states at the hybrid edge and in outer corners are shown.
			Arrangement of spectral charge at (\textbf{d}) ${r<1.0a}$ and (\textbf{e}) ${r>1.0a}$. The red dots denote the position of the Wannier centers. The green unit cells may support states numbered ${91\sim93}$, ${109\sim111}$, ${124\sim126}$, and ${142\sim144}$. The orange unit cells may support states numbered ${112\sim117}$. The blue unit cells may support states numbered ${94\sim99}$, ${118\sim123}$, and ${136\sim141}$. The pink unit cells may support modes numbered $100\sim108$ and $127\sim135$.} 
		\label{fig2}
	\end{figure*}

	The stationary solutions of Eq.~(\ref{eq1}) that propagate in the lattice while maintaining their shapes can be written as ${\psi(x,y,z) = u(x,y) e^{ibz}}$, where $b$ is the propagation constant and $u(x,y)$ is the real-valued function describing the shape of the state. Substitution of the field in such form into Eq.~(\ref{eq1}) yields
	\begin{equation}\label{eq2}
		b u = \frac{1}{2} \nabla^2 u + \mathcal{R}(x, y) u + u^3.
	\end{equation}
	To obtain linear spectrum of the lattice we first neglect the nonlinear term in Eq.~(\ref{eq2}) that transforms it into linear eigenvalue problem. We solve it using plane-wave expansion method and present in Fig.~\ref{fig2}(a) the dependence of propagation constants (eigenvalues) $b$ of all linear modes of the $\mathcal{H}_3$ lattice on shift parameter $r$. The spectrum clearly illustrates two phases, trivial one at ${r<1.0a}$, when the gap opens between two bands of extended bulk states (shown with gray lines), but without any localized states in spectrum, and nontrivial phase at ${r>1.0a}$, where multiple corner (colored lines) and edge states (gray lines) appear in the opened gap. The appearance of latter states is tightly related to the fact that at $r>1.0a$ the intercell coupling becomes stronger than the intracell one driving the system into topological regime. To highlight the structure of eigenmodes in topological regime in Fig.~\ref{fig2}(b) we show eigenvalues, sorted such that $b$ decreases with increase of mode index $n$, versus mode index $n$ at ${r=1.2a}$. In $\mathcal{H}_3$ lattice the modes with indices $n$ from $91$ to $144$ fall into the gap. In accordance with their intensity distributions presented in Fig.~\ref{fig2}(c) (notice that they reflect $\mathcal{C}_3$ discrete rotational symmetry of the lattice) these $54$ modes can be divided into $10$ different groups, as indicated in the figure. Notice that in our continuous model, in contrast to tight-binding models, topological corner states have nonzero propagation constants $b_n$ because we do not use transformation of $\psi$ shifting spectrum by propagation constant of the isolated site. From the intensity distributions shown in Fig.~\ref{fig2}(c), one can see that states numbered ${91\sim93}$, ${109\sim111}$, ${124\sim126}$, and ${142\sim144}$ are localized at the outer three corners of the lattice, while the states numbered ${94\sim99}$, ${112\sim117}$, ${118\sim123}$, and ${136\sim141}$ are localized at different corners at the \textit{hybrid edge} of the lattice and they appear only due to specific structure of the edge. Finally, the states numbered ${100\sim108}$ and ${127\sim135}$ are distributed exclusively at the edges between corners at hybrid edge, so they can be classified as edge states, whose total number is $18$. The states from the same group differ by phases of neighboring groups of spots. Notice that while in the smallest $\mathcal{H}_2$ lattice the states from different corners at hybrid edge can couple, they do not feel each other already in $\mathcal{H}_3$ lattice due to their good localization, achieved already at ${r=1.2a}$ (localization increases with increase of $r$ value), i.e. with increase of the size of the system they turn into isolated, effectively 0D topological corner states. This is the same transition that one observes for conventional corner states in outer corners with increase of the size of the structure. The physical significance of this transition consists in the fact that well-separated states do not couple with each other, so excitation of only one corner on the edge will not lead to coupling of light into other corners on the edge even at distances exceeding sample length by several orders of magnitude.
	
	We would like to stress that in accordance with Fig.~\ref{fig2}(b) the states belonging to groups ${94\sim99}$, ${118\sim123}$, ${136\sim141}$ have distinct eigenvalues and, therefore, the states belonging to these groups are orthogonal. The orthogonality is achieved due to different internal structure of states in different groups. Taking into account that light within corner cells in these states is concentrated mainly on three waveguides, one can show that in states ${94\sim99}$ all three dominating spots in each corner cell are in-phase, in modes ${118\sim123}$ two spots are out-of-phase and there is no light in the waveguide between them due to destructive interference, while in states ${136\sim141}$ two outer spots in corner cells are out-of-phase with central spot. Notice that this phase structure of topological states inside corner cells closely resembles the structure of eigenmodes in three isolated coupled waveguides. Inside the given group of states, e.g. ${94\sim99}$, there is either zero or $\pi$ phase difference between groups of three spots located in different cells and such phase structure is already determined by the geometry of the entire array (it should be consistent with its discrete $\mathcal{C}_3$ rotational symmetry). 
	
	Our simulations show that the structure of the spectrum remains the same in even larger $\mathcal{H}_4$ and $\mathcal{H}_5$ lattices (total number of corner states does not increase in these configurations, while total number of the edge states increases --- for example, in $\mathcal{H}_4$ structure it amounts to $36$). Notice that propagation constants of corner and edge states alternate in gap. 
	In the \textbf{Supporting Information}, we display the spectra of the $\mathcal{H}_2$ and $\mathcal{H}_4$ lattices. One finds that the profile of the spectrum is indeed not affected by the size of the structure.
	It should be also mentioned that linear spectrum of the system presented in Fig.~\ref{fig2}(a) remains qualitatively similar even under strong perturbations that break underlying $\mathcal{C}_3$ rotational symmetry of the structure. In particular, the addition of a waveguide (that should be treated as rather strong perturbation) close to the edge of the structure does not affect most of the corner modes presented here. It only leads to the appearance of a new defect mode localized on newly added waveguide at ${r<1.0a}$ in the gap, and two bulk modes at ${r>1.0a}$ (see the details and mode shapes in \textbf{Supporting Information}).
	
	\subsection{Wannier centers and spectral charges}
	
	The topological origin of the corner states in a finite system can be analyzed via the fractional spectral charge, which is an indication of the ``filling anomaly'' in the language of solid-state physics~\cite{benalcazar.prb.99.245151.2019, li.prb.101.115115.2020, peterson.science.368.1114.2020, peterson.nature.589.376.2021}. The spectral charge can be formally calculated by integrating the local density of all states laying below the band gap that allows to make a conclusion about the number of modes (number of populated sites) per unit cell \cite{liu.nature.589.381.2021}. A more convenient way to extract the spectral charge per unit cell is to consider Wannier centers, which in electronic system with analogous structure would indicate the average position of the electrons~\cite{benalcazar.prb.99.245151.2019}. 
	
	In Fig.~\ref{fig2}(d), we display the distributions of the Wannier centers (red dots) for the lattice in trivial phase, at ${r<1.0a}$. In this case the Wannier centers are located in the centers of the unit cells, hence the spectral charge is $1$ per unit cell. In topological phase, at ${r>1.0a}$, the Wannier centers are known to distribute at the middle of each edge (except for outer edges of those cells that are located at the periphery of the lattice, since the charge fractionalization is a property of the bulk~\cite{benalcazar.prb.99.245151.2019}) of the unit cell, as indicated in Fig.~\ref{fig2}(e). Thus, each Wannier center is shared by two neighboring unit cells, while the spectral charge for each unit cell can be obtained by counting how many Wannier centers are located on the six edges of the unit cell. After normalization to the number of occupied sites in the bulk unit cell~\cite{benalcazar.prb.99.245151.2019, li.prb.101.115115.2020, peterson.science.368.1114.2020, peterson.nature.589.376.2021, shang.as.11.2303222.2024}, one obtains that the spectral charge for the bulk unit cells is integer, while for blue, orange, green, and purple unit cells at the periphery of the lattice it is given by $1/2$, $5/6$, $1/3$, and $2/3$, respectively. This ``filling anomaly'' associated with appearance of fractional charges at ${r>1.0a}$ indicates that corner states localized in corresponding outer cells are of topological origin. In this regard, the lattice in Fig.~\ref{fig1} that supports these corner states can be classified as a crystalline higher-order topological insulator without quantized multipole moments.
	
	
	We would like to point out that the $\mathcal{C}_3$-symmetric HOTI given in Fig.~\ref{fig1} indeed possesses unit cells with variety of possible spectral charges [see Fig.~\ref{fig2}(e)], which are experimentally observable~\cite{liu.nature.589.381.2021}. The topological origin of the corner states can be also analyzed following the method provided in Ref.~\cite{noh.np.12.408.2018} using tight-binding description for photonic graphene lattice with Kekul\'e distortion. This description allows to correctly predict that the phase transition occurs at ${r=1.0a}$ point, but the bulk topological invariant \cite{noh.np.12.408.2018} of the configuration in Fig.~\ref{fig1} in which the corner states are protected by the $\mathcal{C}_6$ rotational symmetry cannot directly distinguish corner states arising due to different fractional charges. Instead, such states connected to different fractional spectral charges can be directly predicted using the Wannier center analysis. On this reason it is more convenient to explain the appearance of different corner states using filling anomaly, as theoretically analyzed in Ref.~\cite{benalcazar.prb.99.245151.2019} and experimentally demonstrated in Refs.~\cite{peterson.nature.589.376.2021, liu.nature.589.381.2021}. To the best of our knowledge, the fractional charge of the unit cell can well predict the appearance of all corner states, which are shown in Figs.~\ref{fig2}(d) and \ref{fig2}(e).
	
	\subsection{Nonlinear topological corner states}
	
\begin{figure*}[t]
		\centering
		\includegraphics[width=\textwidth]{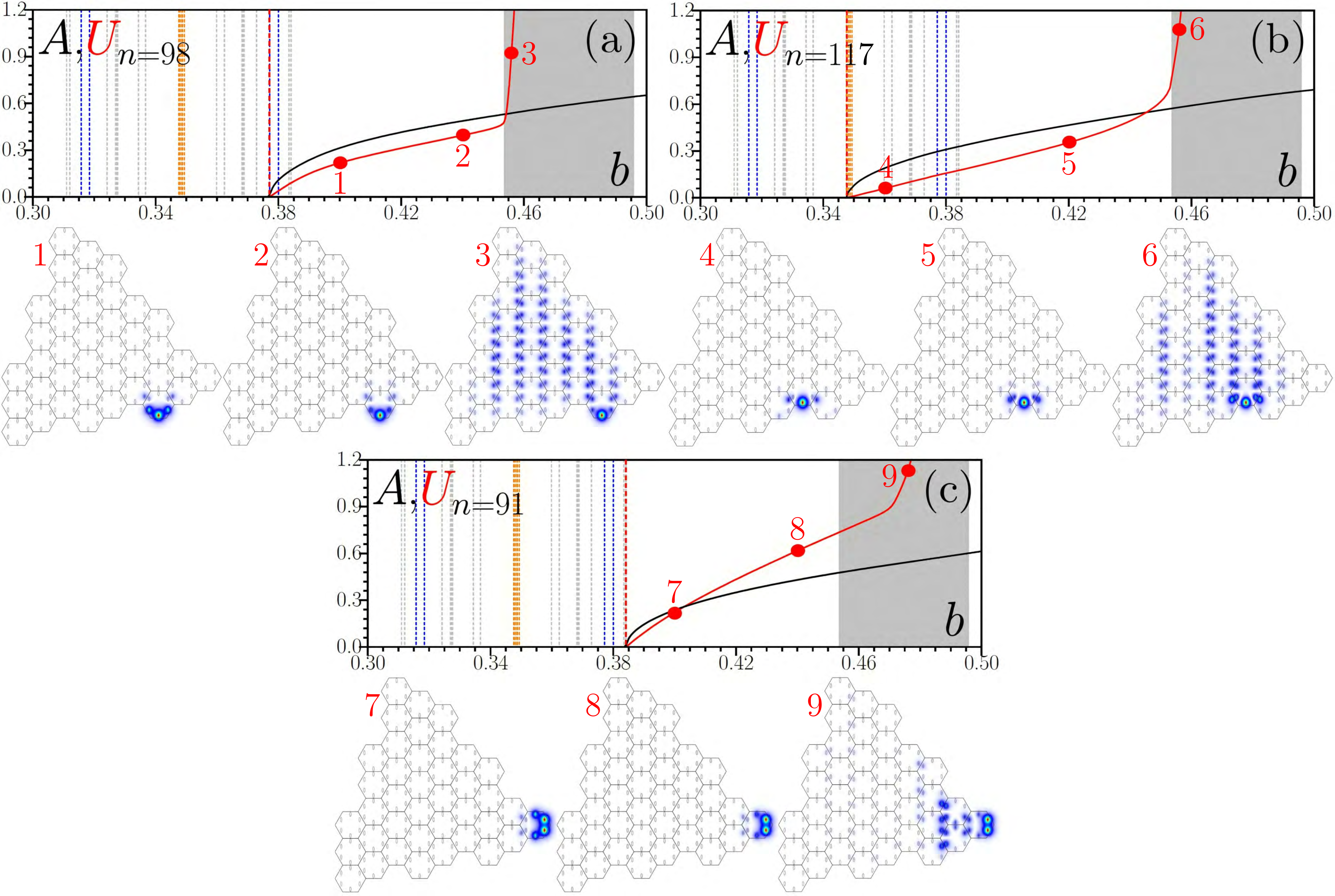}
		\caption{\textbf{Nonlinear topological corner states}. (\textbf{a}) Power $U$ (red curve) and peak amplitude $A$ (black curve) of the nonlinear state bifurcating from the linear state numbered ${98}$. The vertical dashed lines indicate the propagation constants of linear states, and the gray region represents the bulk band. Field modulus distributions correspond to the red dots on the $U(b)$ curve. The unit cells and lattice sites are indicated by gray lines. Panels (\textbf{b},\textbf{c}) show power and amplitude dependencies, but for nonlinear states originating from corner states $117$ and $91$, respectively. In all cases $r=1.2a$.}
		\label{fig3}
	\end{figure*}

	Rich structure of the linear spectrum of the lattice with hybrid edges, where multiple corner states coexist in topological phase, suggests that such linear states can give rise to families of nonlinear corner states when nonlinearity of the medium is taken into account. The nonlinear corner states that we obtain here are the result of the interplay between diffraction, refraction in the inhomogeneous refractive index landscape of the array, and nonlinearity. The interplay between the first two effects creates for a given type of Kekul\'e distortion localized states of topological origin in the system with propagation constants falling into forbidden spectral gap, while nonlinearity strongly affects the shapes (localization) of these states. We obtained such nonlinear states from Eq.~(\ref{eq2}) using the Newton method. At ${r=1.2a}$ linear topological corner states are strongly localized near respective corners (outer ones or at hybrid edges), hence to obtain the family of nonlinear corner states residing in only one desired corner one can use corresponding localized initial guess. For illustrative purposes we present here families originating from corner states numbered $91$, $98$ and $117$, which form in the unit cells with spectral charges $1/3$, $1/2$, and $5/6$, respectively; they include all possible situations for corner cells, according to the Wannier centers analysis. While a variant of state numbered $91$ localized in the outer corner has been observed before, although only in linear case, the other two corner states have never been observed.
	Importantly, different fractional spectral charges, being the result of different configurations of the hybrid lattice border, directly manifest themselves in completely different shapes of the corresponding linear eigenmodes and, consequently, in different shapes of nonlinear modes emerging from them. As we show below, this is also reflected in qualitatively different stability of nonlinear corner state families, as we discuss in Fig. \ref{fig3}. It should be also mentioned that strong nonlinearity may potentially induce a topological phase transition~\cite{maczewsky.science.370.701.2020}, but observation of this scenario requires rather exotic situation, for example, when all inter-cell and intra-cell couplings change differently with intensity of the external pump~\cite{zangeneh.prl.123.053902.2019}. Such effects do not occur in our system because its topological properties are determined by the fabrication process and do not change with increase of pulse energy.
	
	The families of nonlinear topological corner states parameterized by propagation constant $b$ (that in nonlinear problem becomes a free parameter) are depicted in Fig.~\ref{fig3}. Red curves show power of nonlinear corner state ${U=\iint{|u|^2dxdy}}$, while black curves show peak amplitude ${A=\textrm{max}|u|}$. As one can see, nonlinear corner states can bifurcate from different linear corner states numbered $98$ [Fig.~\ref{fig3}(a)], $117$ [Fig.~\ref{fig3}(b)], and $91$ [Fig.~\ref{fig3}(c)]. In the bifurcation point (highlighted by the red dashed line), where ${b\to b_n}$, both amplitude and power of nonlinear corner state vanish, i.e. all such nonlinear corner states are thresholdless. Away from bifurcation point both amplitude and power of nonlinear corner state increase. It is worth stressing that if a linear system possesses forbidden spectral gap with several localized linear states in it, in the presence of nonlinearity each such state can give rise to a family of thresholdless localized nonlinear states that inherit the symmetry of corresponding linear states --- a well-documented fact in theory of nonlinear systems ~\cite{smirnova.apr.7.021306.2020, szameit.np.20.905.2024}. The existence of thresholdless nonlinear corner states illustrated in Fig.~\ref{fig3} is in sharp contrast to properties of lattice solitons in the bulk of the array that can exist only above certain power threshold. Using the theory developed in \cite{konotop.prr.6.033113.2024} one can show that near the linear limit ${b\rightarrow b_n}$, the relation $
		{U(b)=(b-b_n)/\int u_n^4  dxdy}$ is satisfied, with $u_n$ being the profile of corresponding linear state. We have checked that such relations accurately describe $U(b)$ dependencies near the bifurcation point for different corner states, thereby illustrating the connection with shapes of linear localized states.
	
	As one can see from Fig. \ref{fig3}, nonlinearity introduces tunability in this HOTI, as it allows to control the location of the corner state in the gap that is reflected in shapes of corresponding states. In particular, if the propagation constant of the linear topological corner state, from which nonlinear corner state family emerges, is located close to the lower edge of the gap (such states are weakly localized), increasing nonlinearity first causes state shrinkage that is followed by state expansion, when its propagation constant approaches the upper edge of the gap. When the propagation constant is located in the center of the gap, the state is well-localized at low powers and increasing nonlinearity leads to gradual broadening of the state as its propagation constant approaches the upper edge of the gap. Such strong shape transformations of nonlinear corner states clearly indicate on competition between diffraction, refraction, and nonlinearity in the lattice material system and shows that such states are quite far from simple linear corner modes with amplitudes renormalized due to the presence of nonlinear term in Eq.~(\ref{eq1}). In particular, when nonlinearity becomes so strong that propagation constant $b$ enters into the band, nonlinear corner state broadens and then hybridizes with bulk states. This is well visible from the second row in Figs.~\ref{fig3}(a)-\ref{fig3}(c) with exemplary field modulus distributions. Notice that coupling with bulk modes accompanied by increase in the slope of $U(b)$ dependence may occur in the depth of the bulk band and not right at its edge [see Fig.~\ref{fig3}(c)]. Interestingly, as propagation constant $b$ of states $98$ and $117$ increases, it crosses propagation constants of different linear edge states. Nevertheless, no obvious coupling with them occurs for selected nonlinear corner states (due to different symmetries). Situation may be different for other types of nonlinear corner states, not considered here and located closer to the lower edge of the gap, as some of them may couple with edge states. Notice that in trivial phase, when no localized corner modes appear in the gap, the formation of nonlinear corner states in the same spatial locations would occur only above considerable power threshold $U$. Such non-topological nonlinear corner states would be analogous to usual surface solitons with propagation constants in semi-infinite gap of the linear spectrum. Using $U(b)$ dependence one can determine power level at which nonlinear corner state enters into the band that is usually accompanied by a change of slope of this dependence [see Figs.~\ref{fig3}(a) and \ref{fig3}(b)]. Since in experiments the most accessible control parameter is the peak power (defined by the input pulse energy), this allows to estimate when the state will enter into gap also in experiments. Finally, the dependence of peak amplitude $A$ on $b$ highlights that the branch of nonlinear states actually persists even in the band, because $A$ just grows monotonically even when the state penetrates into allowed band.
	
	Notice that the nonlinear corner state families illustrated in Figs.~\ref{fig3}(a) and \ref{fig3}(b) are dynamically stable for propagation constants inside the topological gap (they are unstable when they enter into the bulk band). This implies that nonlinear corner states inherit robustness and topological protection from their linear counterparts from which they bifurcate. This robustness is manifested in the fact that despite complex internal structure of such nonlinear states, the addition of perturbations into underlying optical potential and addition of white noise (up to $10\%$ in amplitude) into input field distributions do not result in destruction of these nonlinear corner states in the entire forbidden gap---they keep propagating in stable fashion maintaining their internal structure, while their peak amplitude exhibits only small oscillations in $z$. This stable evolution is observed in the entire gap, while when the nonlinear corner state enters into the band, it may start radiating due to coupling with bulk states. In contrast, the nonlinear corner state family with different symmetry associated with cell with different fractional spectral charge and shown in Fig.~\ref{fig3}(c) is stable only close to the bifurcation point from linear corner state, and becomes unstable within the gap, around ${b\sim0.43}$. This is the consequence of symmetry of this particular solution with two equally strong in-phase spots in two waveguides that makes it prone to symmetry-breaking instability. The details of stability analysis for nonlinear corner state families can be found in \textbf{Materials and Methods} section and \textbf{Supporting Information}.
	
	\subsection{Experimental observation of the nonlinear corner states}
	
	To demonstrate linear topological corner states and nonlinear corner states in lattice with hybrid edges, we fabricated in $10~\textrm{cm}$-long fused silica slabs using the fs-laser direct writing technique (details of fabrication procedure can be found in the \textbf{Materials and Methods} section, while the scheme of the experimental setup is presented in \textbf{Supporting Information}) a set of waveguide lattices with various waveguide shifts $r$ in the unit cells, including configurations depicted in Figs.~\ref{fig1}(c)-\ref{fig1}(e). Micro-photographs of exemplary lattices presented in Fig.~\ref{fig1} illustrate the system in trivial phase with ${r=0.8a}$ [Fig.~\ref{fig1}(f)], the borderline setting with ${r=1.0a}$ [Fig.~\ref{fig1}(g)], and the lattice in topological phase with ${r=1.2a}$ [Fig.~\ref{fig1}(h)]. The methods to control the accuracy of fabrication of the waveguide arrays and reproducibility of the experimental results are discussed in the \textbf{Materials and Methods} section. Notice that the efficiency of excitation of different states is determined exclusively by the \textit{overlap integral} ${\mathcal{O}_n=\int u_n^* \left.\psi\right|_{z=0} dxdy}$ of the input beam $\psi$ at ${z=0}$ with modal shape $u_n$ that depends on two main factors: (i) the proximity of the beam to the cell, where the eigenstate resides, and (ii) the parity of the input beam and of the state. Thus, for efficient excitation (yielding maximal $\mathcal{O}_n$) of states ${112\sim117}$ or ${94\sim99}$ a single input Gaussian beam focused into waveguide (site 2 or 1) where state maximum resides, is sufficient as it provides maximal overlap with these states, but for excitation of corner modes ${91\sim93}$ or ${142\sim144}$ one should use in-phase or out-of-phase combination of two beams, respectively, due to different parities of these states. Accordingly, when the input beam is selected such that it simultaneously excites two states, one may observe beating between them in linear case that can be arrested by the nonlinearity. Taking into account the above considerations we use single-site excitations of waveguides 1 and 2 denoted on micro-photographs with orange and blue circles (as they would have largest overlap with modes $98$ and $117$ discussed above), and in-phase excitation of two outer corner waveguides 3 denoted with green ellipse (this would create an input with largest overlap with mode $91$). To achieve high peak intensities, at which nonlinearity of fused silica starts affecting propagation of light in inhomogeneous refractive index landscape, we use for excitation pulses of variable energy $E$ from ${1\,\rm kHz}$ Ti:sapphire laser at ${800\,\rm nm}$ central wavelength focused into selected waveguides. Sufficiently large pulse duration of $280~\textrm{fs}$ ensures that dispersion broadening of pulses on $10~\textrm{cm}$ sample length can be neglected, therefore our purely spatial model adequately describes propagation dynamics (see \textbf{Supporting Information} for discussion of the role of temporal effects). In the case of two-site excitations we employed a Michelson interferometer to separate laser beam into two. Two compensation plates in one arm of the interferometer were used to smoothly control the relative phase of the beams without undesirable spatial displacement. A detailed description of the experimental setting can be found in the \textbf{Materials and Methods} section and \textbf{Supporting Information}.
	
\begin{figure}[t]
		\centering
		\includegraphics[width=0.5\columnwidth]{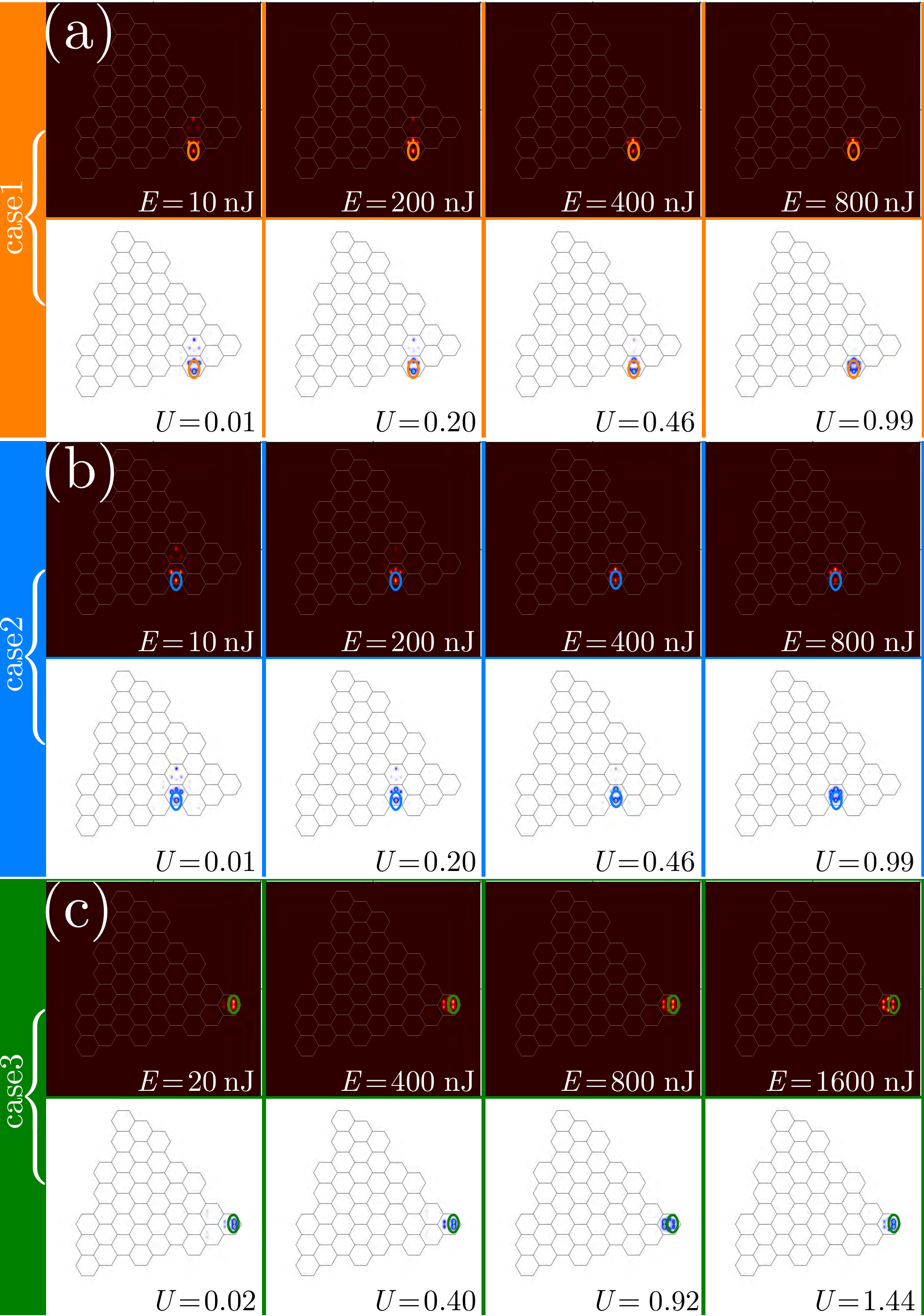}
		\caption{\textbf{Excitation of nonlinear states in trivial phase, at ${r=1.0a}$}. (\textbf{a}) Output intensity distributions for excitation of site 1 at hybrid edge indicated by the orange circle here and in micro-photograph in Fig. \ref{fig1}(g) (in topological regime this excitation would have largest overlap with state numbered ${n=98}$). (\textbf{b}) shows output intensity distributions for excitation of site 2 indicated by the blue circle that would produce largest overlap with state numbered ${n=117}$, while (\textbf{c}) shows in-phase excitation of two sites indicated by the green circle [see also Fig. \ref{fig1}(g)] that has strongest overlap with outer corner state numbered ${n=91}$. The panels with maroon and white background show experimental and theoretical results, respectively. The input powers (pulse energies in experimental images) are indicated for each panel.}
		\label{fig4}
	\end{figure}

	First we consider the excitation dynamics in borderline situation, at ${r=1.0a}$, see Fig.~\ref{fig4}. In this figure experimental output intensity distributions (images with maroon background) are compared with results of theoretical simulations in Eq.~(\ref{eq1}) (images with white background). Since at ${r=1.0a}$ all states of the lattice are delocalized (because intercell and intracell coupling strengths are identical in the entire lattice), excitation of any corner at hybrid edge [Figs.~\ref{fig4}(a) and \ref{fig4}(b)] or outer corner [Fig.~\ref{fig4}(c)] results in gradual beam diffraction both in the linear regime (pulse energy ${E=10\,\rm nJ}$) and in weakly nonlinear regime (pulse energy ${E=200\,\rm nJ}$). This is because such input excites multiple delocalized states of the lattice with comparable weights. Light redistributes between all waveguides in the excited cell and even penetrates into neighboring cells. Notice that in Fig.~\ref{fig4}(c) concentration of light in two corner waveguides is a result of dynamical beatings and not an indication of stationary corner state formation (this is confirmed by the fact that increasing input power leads to escape of light from these two waveguides, see intensity distribution for ${E=800\,\rm nJ}$). The diffraction gradually slows down with increase of pulse energy with a tendency to contraction to the excited waveguide, but even at pulse energies of ${E=800\,\rm nJ}$ close to the damage threshold of the material (for excitation in Fig.~\ref{fig4}(c) we use ${E=1600\,\rm nJ}$ to reflect the fact that two waveguides are simultaneously excited), we do not observe yet the concentration of light in a single waveguide (i.e. the power is insufficient to excite usual non-topological surface soliton in this regime).
	
\begin{figure}[t]
		\centering
		\includegraphics[width=0.5\columnwidth]{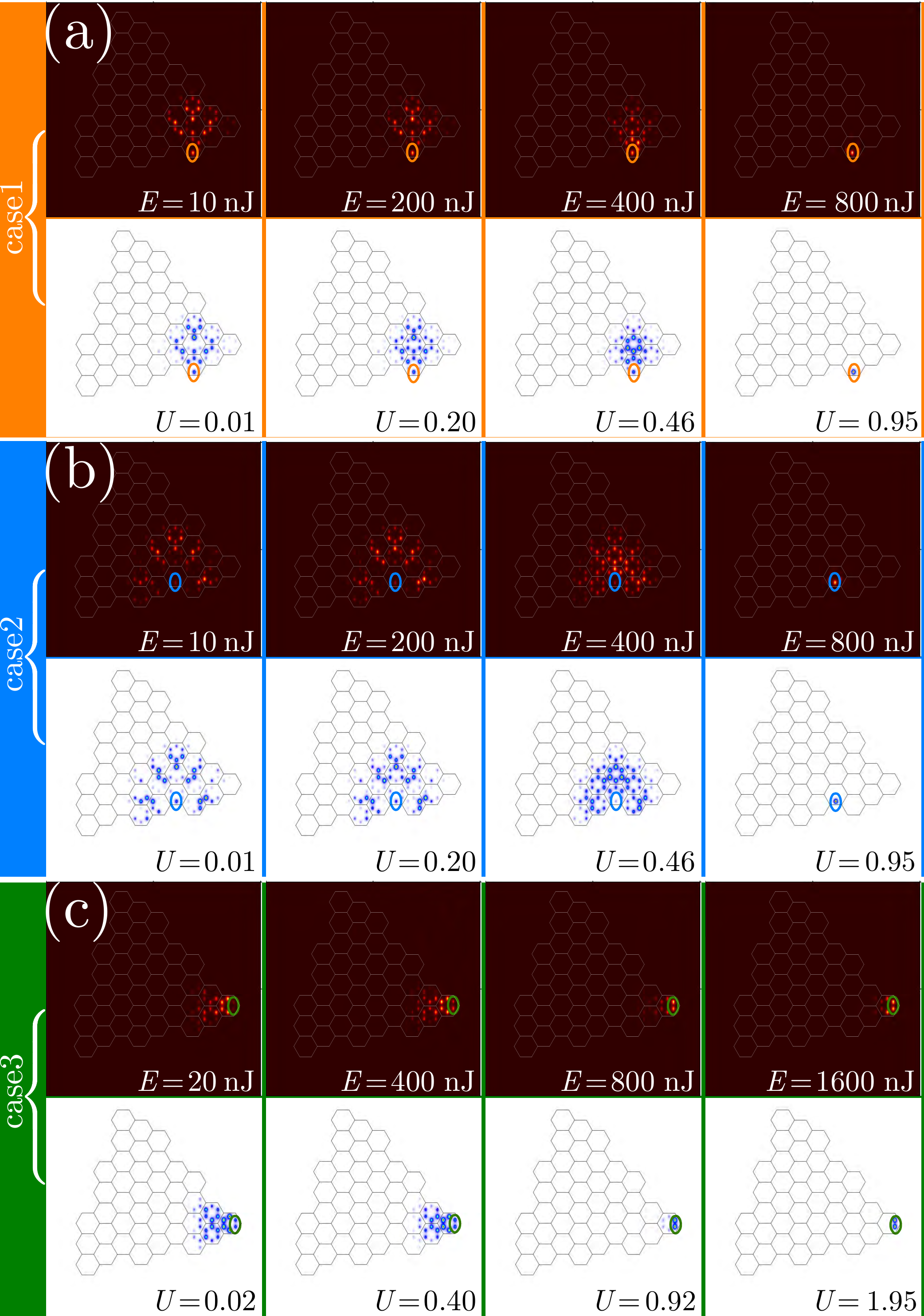}
		\caption{\textbf{Excitation of nonlinear states in trivial phase, at ${r=0.8a}$}. The arrangement of panels and notations are the same as in Fig.~\ref{fig4}, but the regime deeper in trivial phase is selected.}
		\label{fig5}
	\end{figure}

	When we excite the lattice with shift parameter ${r=0.8a}$ deeper in the trivial regime [see Fig.~\ref{fig1}(f)], one observes dramatic enhancement of diffraction in linear regime, as illustrated in Fig.~\ref{fig5}. It is observed for all types of low-power excitations, both at hybrid edge [Figs.~\ref{fig5}(a) and \ref{fig5}(b)] and in the outer corner [Fig.~\ref{fig5}(c)]. However with increase of input pulse energy up to ${E=800\,\rm nJ}$ the sharp transition to nonlinear localization in single site is observed, and in the case of two-site excitation one also sees the onset of localization at ${E=1600\,\rm nJ}$. This is a clear indication of formation of \textit{non-topological} surface solitons above certain power threshold. Apparently, such solitons have propagation constants in semi-infinite gap, laying above upper bulk band in Fig.~\ref{fig2}(a). A somewhat surprising observation is that power threshold for formation, which is a typical feature of such surface solitons, appears to be lower at ${r=0.8a}$ in comparison with ${r=1.0a}$ lattice.
	
	\begin{figure}[t]
		\centering
		\includegraphics[width=0.5\columnwidth]{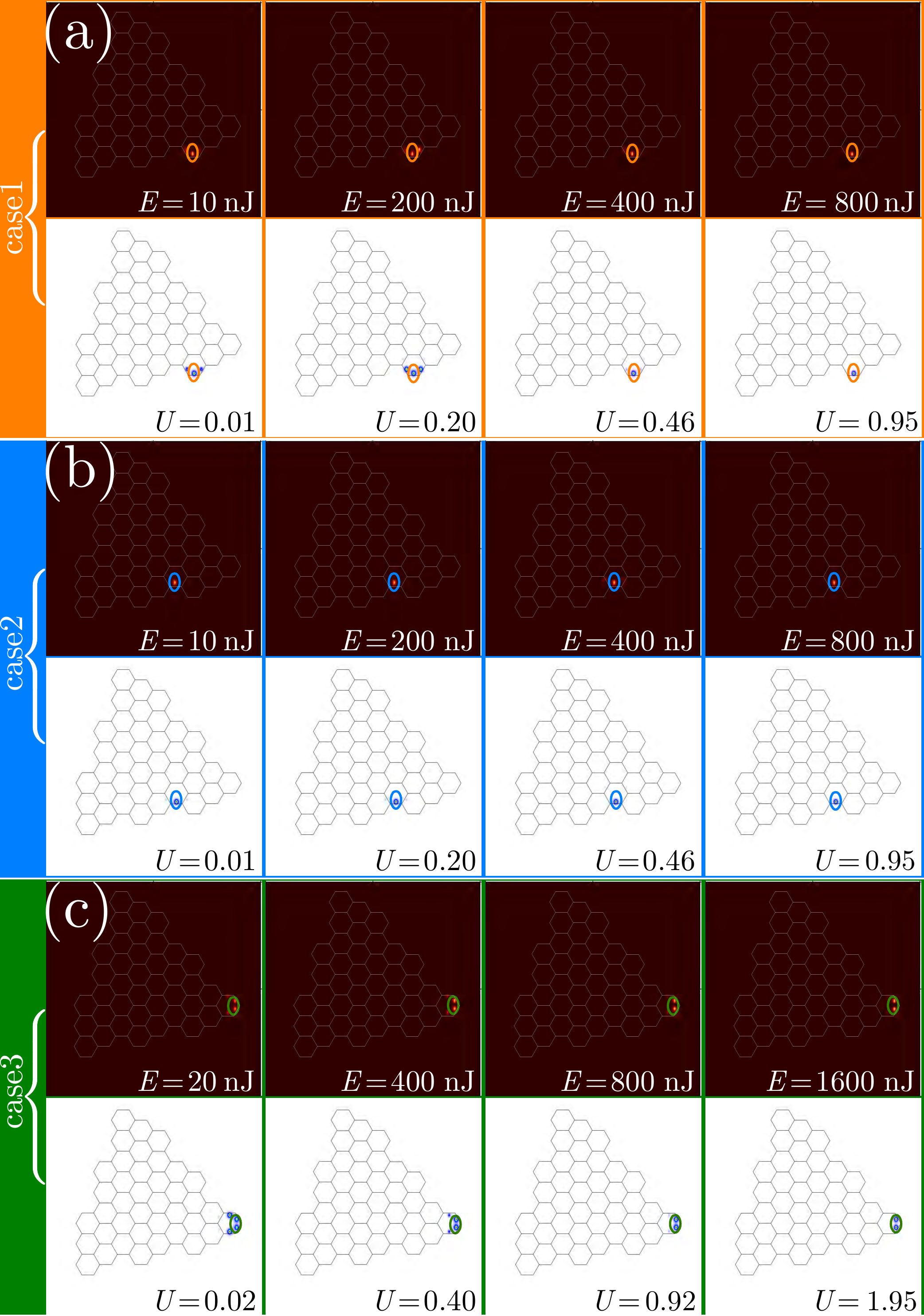}
		\caption{\textbf{Excitation of nonlinear topological corner states at ${r=1.2a}$}. The arrangement of panels is the same as in Fig.~\ref{fig4}, but now the lattice in topological regime is considered.}
		\label{fig6}
	\end{figure}
	
	The excitation of lattice in topological regime, with ${r=1.2a}$, whose photograph is depicted in Fig.~\ref{fig1}(h), leads to sharply distinct dynamics. In Fig.~\ref{fig6}(a) we illustrate the excitation of site 1, and in clear contrast to previously considered lattices light remains strongly confined around the excited corner waveguide at hybrid edge even in the linear regime at ${E=10\,\rm nJ}$, only slightly penetrating into two nearest neighbors (see experimental results in images with maroon background). This intensity distribution closely resembles the profile of the nonlinear corner state near the bifurcation point from the corner state numbered ${n=98}$, see Fig.~\ref{fig3}(a). The localization of this state further improves with increase of input pulse energy, see distributions for ${E=200\sim800\,\rm nJ}$ indicating on the formation of nonlinear topological corner state at hybrid edge. Notice that in agreement with nonlinear corner state family presented in Fig.~\ref{fig3}(a), this state does not demonstrate coupling with edge states. Coupling with bulk modes also was not observed for this and for other types of corner excitations. One of the reasons is that such coupling occurs in relatively narrow range of input powers for selected $r$ value. The second reason is that even though our $10~\textrm{cm}$ sample is long and allows to observe rich nonlinear dynamics, coupling with bulk modes is a slow process that requires distances that may substantially exceed the length of the sample. When site 2 is excited, we observe even at low pulse energies of ${E=10\,\rm nJ}$ the excitation of nonlinear corner state bifurcating from state numbered ${n=117}$ [see Fig.~\ref{fig3}(b)] that shows good localization up to the pulse energies ${E=800\,\rm nJ}$, as shown in Fig.~\ref{fig6}(b). It should be stressed that already in $\mathcal{H}_3$ lattice, even in linear regime, single-site corner excitations at hybrid edge do not couple into similar corners along the edge, even at the distance exceeding sample length by an order of magnitude, which means that the excited states are indeed isolated 0D states. Finally, excitation of nonlinear corner states in outer corner 3 with two-site in-phase input (corresponding to the theoretical results in Fig.~\ref{fig3}(c) illustrating bifurcation from state numbered ${n=91}$) is illustrated in Fig.~\ref{fig6}(c). In this case, one also observes gradually increasing with input pulse energy localization inside two corner waveguides. Results of theoretical simulations illustrated in images with white background agree well with experimental results.
	
	To confirm that the corner states reported here are indeed nonlinear objects, we note that in topologically trivial structures (particularly with ${r=0.8a}$ in Fig.~\ref{fig5}), where one observes considerable diffraction broadening and displacement of the output pattern into the bulk at low powers, the impact of nonlinearity is obvious, since increasing power leads to gradual localization of the output. This means that nonlinearity-induced change of refractive index at highest input pulse energies was strong enough to localize light practically in a single waveguide even in nontopological regime. Since in topological regime at ${r=1.2a}$ in Fig.~\ref{fig6} we employ the same pulse energies, the same levels of nonlinear correction to the refractive index are reached. However, since transformations of the output spatial intensity distributions at different pulse energies are not so pronounced at ${r=1.2a}$, we also measured the evolution of the output pulse spectrum in the center of the excited waveguide (for site 2 excitation) with increase of input pulse energy. This evolution is presented in \textbf{Supporting Information}. Considerable broadening of the spectrum due to self-action effects with increase of pulse energy to ${800\,\rm nJ}$ was observed (that starts already at pulse energies around $400\,\rm nJ$). This clearly indicates that the observed topological corner states propagate in nonlinear regime.
	
	\section{Conclusions}
	Linear corner states and nonlinear corner states of diverse types, observed here in HOTI with hybrid edges and $\mathcal{C}_3$ discrete rotational symmetry, illustrate the role of the edge geometry and internal structure of HOTI in defining the complexity and richness of the spectrum of modes arising in topological phase. We have shown that all types of in-gap corner states arise due to filling anomaly, which leads to the fractional spectral charge (including $1/3$, $1/2$, and $5/6$) on the corresponding unit cells. When such lattices are created in nonlinear medium, enriched spectrum of topological modes gives rise to multiple families of nonlinear topological corner states arising in different corners along the hybrid edge. Our results open new avenues for control of light beam propagation and localization. The diversity of nonlinear corner states appearing in our system is quite large --- they emerge with different shapes depending on the spectral charge of corresponding cell and in different spatial locations --- and this may be advantageous for design of topological corner lasers with different frequencies and topological rainbow devices \cite{kim.nc.11.5758.2020, zhang.light.9.109.2020, zhong.apl.6.040802.2021}. In addition, the results also help to deeply understand the higher-order band topology of lattice materials that may inspire applicative schemes in fabricating on-chip optical functional devices.
	
	Topological materials with hybrid edges and corner states originating from different spectral charges may open rich prospects for the design of novel photonic devices and may be useful for various practical applications~\cite{lu.chip.1.100025.2022}. Such corner states can be used for the design of transmission devices for signals that are not affected by disorder in the underlying structure due to the topological nature of the states~\cite{peterson.nature.555.346.2018,jiang.rip.13.100112.2025}. 
		This applies not only to materials with purely spatial settings, where the information can be encoded and transmitted in topological states with \textit{different} internal spatial structure residing in \textit{different} corners, but also to spatiotemporal situations~\cite{ivanov.pra.107.033514.2023} where light pulses can be transmitted in ``topological waveguides'' originating in different corners. 
		In this case, the spatial structure of corner state (which is very different for different spectral charges) in the nonlinear regime defines the effective mode area and the effective nonlinear coefficient that determines the \textit{temporal} dynamics of light pulses (or interaction of several such pulses) transmitted in \textit{different} corner states. 
		Structured materials with hybrid edges can be used for the design of topological lasers~\cite{ota.nano.9.547.2020,zhang.light.9.109.2020,zhong.apl.6.040802.2021}, where lasing can be achieved in corner states with different internal structure associated with different spectral charges, depending on the pump configuration that defines the lasing thresholds for the corner states. 
		For materials with hybrid edges, these thresholds (and the slope efficiency) may be different, unlike in conventional HOTIs with identical corners. 
		Such lasers can be built on waveguiding structures inscribed in active materials~\cite{dupont.ol.50.666.2025}, and also in polariton microcavities~\cite{klembt.nature.562.552.2018,kartashov.prl.122.083902.2019}, where microcavity patterning technologies are well developed and rich possibilities for designing pump landscapes (to control lasing thresholds) are reported. 
		A notable feature of polariton systems is the possibility of using a resonant pump that allows selective resonant excitation (lasing) of corner states with different eigenfrequencies, when the latter match the pump frequency. 
		In material systems with hybrid edges, such eigenfrequencies are different for corners with different spectral charges, which makes it possible to achieve lasing only in a \textit{desired subset of corners} and switch to lasing in another subset of corners only by changing pump frequency, even with uniform pump. 
		In this respect, the material system with hybrid edges and different types of corners is advantageous compared to conventional HOTI because it enriches possible lasing regimes. Another potential application of materials supporting corner states originating from different spectral charges is the realization of quantum memory~\cite{heshami.jmo.63.2005.2016,moiseev.prl.134.070803.2025,agarwal.prb.107.125163.2023}. In this case, the recording of the information can be done by resonant absorption of the topological states excited in different corners of the material, while the reading of the information can be initiated by the control pulses leading to the stimulated emission of light in the topological state in the same corner, preserving all the information about the structure of the state. 
		In this case, the number of different corners determines the amount of information that can be recorded, while the topological nature of the states makes the material system resistant to errors. 
		This is another area where coexistence of \textit{several different types} of topological states is advantageous compared to conventional HOTIs.
	
	\section*{Materials and Methods}
	
	\subsection*{Normalization of parameters in theoretical model}
	
	The dimensionless Eq.~(\ref{eq1}) in the main text is derived from the following dimensional version:
	\begin{equation}\label{eq3}
		i \frac{\partial \mathcal{E}}{\partial \zeta} = -\frac{1}{2k} \left( \frac{\partial^2}{\partial \xi^2} + \frac{\partial^2}{\partial \eta^2} \right) \mathcal{E} - \frac{k}{n_0}(\delta n + n_2|\mathcal{E}|^2) \mathcal{E},
	\end{equation}
	where $\xi,\eta$ are transverse coordinates, and $\zeta$ is the propagation distance. We use standard ``soliton'' units and introduce dimensionless transverse coordinates as ${x=\xi/r_0}$, ${y=\eta/r_0}$, where ${r_0=10~\mu \textrm{m}}$ is the characteristic transverse scale, that also determines the relation between the real propagation distance $\zeta$ and the dimensionless propagation distance $z$ as ${\zeta=z_d z}$, where the diffraction length ${z_d=kr_0^2\approx1.14\,\textrm{mm}}$. Real field amplitude is related to dimensionless one as ${\mathcal{E}=(n_0/k^2 r_0^2 n_2)^{1/2} \psi}$. Here ${k=2\pi n_0/\lambda}$ is the wavenumber in the medium with unperturbed refractive index $n_0$ (for fused silica ${n\approx 1.45}$ and the nonlinear refractive index ${n_2\approx 2.7\times 10^{-20}\, \rm m^2/W}$), ${\lambda=800~\textrm{nm}}$ is the experimental working wavelength, and $\delta n$ is the refractive index contrast defining the structure of shallow optical potential $\mathcal{R}(x,y)$. The dimensionless depth of this potential is given by ${p=k^2r_0^2\delta n/n_0}=4.7$ that corresponds to the refractive index contrast ${\delta n \approx 5.3\times 10^{-4}}$. The side length of the unit cell of the lattice in all considered structures is ${a=3.3}$ that corresponds to ${33~\mu\textrm{m}}$, waveguide widths ${d_x=0.25}$, ${d_y=0.75}$ correspond to ${2.5~\mu\textrm{m} \times 7.5~\mu\textrm{m}}$ wide elliptical waveguides, while sample length of ${10~\textrm{cm}}$ corresponds to ${z\sim 88}$.
	
	Since the experiments are performed at ${\lambda=800\,\rm nm}$ wavelength, i.e. in normal dispersion regime where ${k''=\partial^2k/\partial \omega^2 \approx 36.163\,\rm fs^2/mm}$ in fused silica, the localization of pulses in time is impossible in this system with focusing nonlinearity. It is on this reason, experimental studies of topological solitons or nonlinear effects in fs-laser written arrays performed in this spectral range used sufficiently long pulses for which temporal dynamics can be disregarded and only spatial output intensity distributions may be controlled (see, for example, Refs.~\cite{xia.light.9.147.2020, kirsch.np.17.995.2021, mukherjee.prx.11.041057.2021, hu.light.10.164.2021}). This is the case also for our experiments. Nevertheless, to prove that temporal effects are not essential in our system, we modeled excitation dynamics using (3+1)D extension of Eq.~(\ref{eq1}). The results show that averaged output intensity distributions obtained in (3+1)D case are practically indistinguishable from output intensity distributions in purely spatial model. For modeling detailed discussion see \textbf{Supporting Information}.
	
	\subsection*{Fs-laser inscription of the waveguide arrays in fused silica}
	Honeycomb lattices with hybrid boundaries were written in ${10\,\textrm{cm}}$ long fused silica substrate (JGS1). Femtosecond pulses with a central wavelength of $515\,\textrm{nm}$, pulse duration of ${230\,\textrm{fs}}$, repetition rate of ${1\,\textrm{MHz}}$, pulse energy of ${320\,\textrm{nJ}}$ and circular polarization were focused by an aspheric lens (${\textrm{NA} = 0.3}$) into the bulk of the substrate. Glass sample was translated with respect to the beam waist at a scanning velocity of ${1\,\textrm{mm/s}}$ by a high-precision positioner (AeroTech FiberGlide 3D). Fabrication of almost depth independent waveguides in the depth range from $400$ to $1000\,\mu\textrm{m}$ was provided by soft focusing. Waveguides demonstrate propagation losses below ${0.3\,\textrm{dB/cm}}$ at ${\lambda = 800\,\textrm{nm}}$.
	
	Fs-laser writing is a powerful, stable, and well-developed technology allowing inscription of waveguide arrays with highly reproducible parameters and minimal variations in positions of the waveguides, guaranteed by the high-precision positioning system Aerotech used for sample translation upon waveguide inscription~\cite{skryabin.prap.22.064079.2024}. Refractive index contrast in waveguides is determined by writing laser parameters, and some small fluctuations are, of course, unavoidable because pulse duration stability of writing laser may be affected by environmental changes despite the use of power stability system, especially during long writing process of series of waveguide arrays with different $r$ parameters. To control reproducibility of waveguides in the array we use auxiliary identical 1D or 2D arrays with equally spaced waveguides written between arrays with hybrid edges. Such auxiliary arrays demonstrate exceptionally symmetric discrete diffraction patterns, when central waveguide is excited, that are practically identical. This allows to exclude the following two main sources of perturbations:
		\begin{enumerate}
			\item[(1)] Short-term laser instabilities that could lead to differences in depths of neighboring waveguides in the \textit{same} array. In arrays with equally spaced waveguides such differences would disrupt the symmetry of the discrete diffraction pattern, so if the latter remain symmetric the waveguide parameters can be considered stable at the timescale of writing of single array.
			\item[(2)] Environmental fluctuations that occur at large timescales needed for writing of several arrays (almost a day in the current work) that could lead to differences in depths of waveguides in \textit{different} arrays written with large time interval. To exclude such variations, we compare the rates of discrete diffraction in auxiliary arrays written between arrays with hybrid edges to make sure they are identical.
	\end{enumerate}
	
	This confirms that the experimental results are fully reproducible. Using auxiliary 1D or 2D arrays of equally spaced waveguides allows not only to monitor short- and long-term reproducibility of waveguides in the entire sample, but also to reproduce the structures with identical coupling strengths between waveguides even after long time periods and for different energies of writing pulses and depths of writing.
	
	\subsection*{Experimental excitation of the waveguide arrays}
	
	In our experiments, we used femtosecond pulses from a Ti:sapphire laser system, Spitfire HP, with a central wavelength of ${800\, \rm nm}$ to excite the waveguides. The pulse repetition rate was ${1\,\rm kHz}$. Short, ${40\,\rm fs}$, broad-spectrum pulses from the regenerative amplifier system first pass through an active beam position stabilization system that minimizes slow spatial fluctuations of the beam, and are then fed into a $4f$ reflective zero-dispersion compressor that acts as a pulse slicer, cutting out a narrow spectrum with a half-width of ${5\,\rm nm}$. A detailed schematic of the experimental setup and the pulse slicer is provided in the \textbf{Supporting Information}. The pulse duration with this spectral width at the output was ${280\,\rm fs}$. This increase in pulse duration helps to avoid too strong spectral broadening during pulse propagation in the waveguides, allowing temporal effects to be neglected. To investigate the influence of spectral width on nonlinear processes, we conducted experiments with an even narrower spectrum of ${2\,\rm nm}$ and a pulse duration of ${700\,\rm fs}$, which fully replicated the nonlinear dynamics of radiation propagation. After the pulse slicer, the radiation entered a Michelson interferometer, necessary for two-site excitations. In the interferometer, a pair of compensation plates, set at the Brewster angle, was used for precise phase control. Simultaneous rotation of these plates in opposite directions allowed for accurate control of the phase of the two pulses without altering the direction of the radiation. For single-site excitation experiments, one arm of the interferometer was blocked. After the interferometer, the radiation was focused onto the input facet of the sample using an aspheric lens. In our experiments, we used a lens with a focal length of ${100\,\rm mm}$, which, for a beam diameter of ${6\,\rm mm}$, provided optimal coupling with the waveguide modes. The calculated overlap integral for this lens and the elliptical mode of the waveguide was 0.847, resulting in a theoretical coupling loss of ${0.72\,\rm dB}$. Accounting for Fresnel reflections and coupling losses, we estimate that each ${1\,\rm nJ}$ in the pulse corresponds to ${2.5\,\rm kW}$ of power in the waveguide for a ${280\,\rm fs}$ pulse duration. To align the input radiation with the waveguides, a precision 6-axis I6000 XYZ/RYP Positioner (XY/4-10 nm) from Luminos was used, enabling sample movement with nanometer accuracy. The intensity distributions at the output of the waveguide arrays were magnified by a factor of 4 and recorded using a 12.3 MP CMOS scientific camera.
	
	We also checked that excitation of similar waveguides at different boundaries of our arrays yield similar diffraction patterns (for a given $r$) that are only slightly affected by the ellipticity of the waveguides and that match each other very well after proper rotation of the pattern by the angle ${\sim2\pi/3}$ in accordance with discrete rotational symmetry of the structure. We also checked that series of excitations of the same site (for example, site 1) in arrays with ${r=0.8a}$, $1.0a$, $1.2a$ yield at low input power nearly identical (for a fixed $r$) diffraction patterns with very close values of the form-factor ${\chi=(U^{-2} \int I^2 dxdy)^{1/2}}$, where ${U=\int Idxdy}$, that is inversely proportional to the width of the diffraction pattern and that rapidly increases with increase of $r$, reflecting the formation of localized states at ${r>1.0a}$. This confirms reproducibility of output patterns for a given $r$ and excitation position from measurement to measurement.
	
	\subsection*{Linear stability analysis of the nonlinear corner states}
	
	We perturb the nonlinear corner state solutions with perturbations $v(x,y)$ and $w(x,y)$. Their growth rate is labeled as $\beta$, and their amplitude is much smaller than the amplitude of the nonlinear corner state solution $u(x,y)$. Therefore, the perturbed solution of Eq.~(\ref{eq1}) can be written as
	\begin{equation}\label{eq4}
		\psi = \left[ u(x, y)+v(x, y) e^{\beta z}+w^{*}(x, y) e^{\beta^{*} z} \right] e^{i b z},
	\end{equation}
	Clearly, if ${\beta_{\rm re}={\rm Re}\{\beta\}\le0}$ for all perturbation modes, then nonlinear corner states are stable, otherwise if at least one perturbation with ${\beta_{\rm re}>0}$ exists, they are unstable. Plugging the perturbed solution in Eq.~(\ref{eq4}) into Eq.~(\ref{eq1}), and linearizing it around stationary solution $u(x,y)$, we obtain the linear eigenvalue problem:
	\begin{equation}\label{eq5}
		\beta
		\begin{bmatrix}
			v\\
			w
		\end{bmatrix}
		=
		\mathcal{M}
		\begin{bmatrix}
			v\\
			w
		\end{bmatrix}.
	\end{equation}
	Here the matrix
	\[
	\mathcal{M} = -|u|^2 \sigma_2 + i \left( \frac{1}{2}\nabla^2 + {\mathcal R}-b + 2|u|^{2} \right)\sigma_3
	\]
	is constructed using nonlinear corner state solution $u$, and $\sigma_{2,3}$ are the Pauli matrices. By solving Eq.~(\ref{eq5}) using standard eigenvalue solver we obtain the dependencies $\beta(b)$ for perturbation modes. The numerical results (see details in the \textbf{Supporting Information}) demonstrate that nonlinear corner state families illustrated in Figs.~\ref{fig3}(a) and \ref{fig3}(b) are stable inside the topological gap, while the family shown in Fig.~\ref{fig3}(c) changes its stability with increase of peak amplitude of solution.

	\medskip
	\noindent \textbf{Supporting Information} \par 
	\noindent Supporting Information is available from the Wiley Online Library or from the author.
	
	\medskip
	\noindent \textbf{Data availability} \par
	\noindent The data that support the plots within this paper and other findings of this study are available from the corresponding author on reasonable request.
	
	\medskip
	\noindent \textbf{Code availability} \par
	\noindent The analysis codes will be made available on reasonable request.
	
	\medskip
	\noindent \textbf{Author contributions} \par
	\noindent Y. Q. Zhang and Y. V. Kartashov formulated the problem. S. A. Zhuravitskii, N. N. Skryabin, I. V. Dyakonov, and A. A. Kalinkin fabricated the samples. V. O. Kompanets, A. V. Kireev and S. V. Chekalin performed experiments. S. Feng, Y. Q. Zhang, Y. D. Li, and C. Shang performed numerical modeling. Y. V. Kartashov, S. P. Kulik, and V. N. Zadkov supervised the work. All co-authors took part in discussion of the results and writing manuscript.
	
	\medskip
	\noindent \textbf{Declaration of Interests} \par
	\noindent The authors declare no competing interests.
	
	\medskip
	\noindent \textbf{Acknowledgements} \par 
	\noindent This work was supported by the Natural Science Basic Research Program of Shaanxi Province (2024JC-JCQN-06), the National Natural Science Foundation of China (12474337), the research project FFUU-2024-0003 of the Institute of Spectroscopy of the Russian Academy of Sciences and the Russian Science Foundation (24-12-00167). S. A. Zhuravitskii acknowledges support by the Foundation for the Advancement of Theoretical Physics and Mathematics ``BASIS'' (22-2-2-26-1).
	
	\medskip
	
	%
	
	%
	
\end{document}